\documentclass[preprint,showpacs, aps]{revtex4}
\usepackage{graphicx}
\usepackage{amsmath}
\usepackage{amssymb}

\begin{document}

\title{Magnetotransport through graphene spin valves}

\author {Kai-He Ding$^1$, Zhen-Gang Zhu$^2$, and Jamal Berakdar$^2$}
\affiliation{$^1$Department of Physics and Electronic Science,
Changsha University of Science and Technology, Changsha,410076,
China\\ $^2$Institut f\"{u}r Physik Martin-Luther-Universit\"{a}t
Halle-Wittenberg Nanotechnikum-Weinberg,
 Heinrich-Damerow-Strasse 4 D - 06120 Halle (Saale), Germany}


 \begin{abstract}
 We present a theoretical study on the spin-dependent transport through a spin
 valve consisting of graphene sandwiched between two magnetic leads
 with an arbitrary orientation of the  lead magnetization.
 No  gate voltage is applied. Using Keldysh's nonequilibrium Green's function method we show that,
 in absence of external magnetic fields, the current-voltage curves are  nonlinear.
 Around  zero bias the  differential conductance versus bias voltage possesses a strong dip.
 The zero-bias anomaly  in the tunnel magnetoresistance (TMR) is affected strongly
 by the leads spin polarization. Depending on the value of the  bias voltage TMR exhibits
 a behavior ranging from  an insulating to a metallic-type. In presence of a static external
 magnetic field the differential conductance and TMR as a function of the bias
 voltage and the strength of the magnetic field show periodic oscillations due to Landau-level crossings.
 We also inspect the effects of the temperature and the polarization degrees on the
 differential
 conductance and TMR.
 \end{abstract}


 \maketitle%

 \section{Introduction}
 Recent advances in nanoscience
 techniques opened the way for  the creation and the investigation of the two-dimensional carbon,
 also called graphene \cite{novoselov1}\cite{Novoselov2005pnas}\cite{zhang2005prl}\cite{berger2004jpcb}.
 This system is a monolayer of carbon atoms packed densely into a honeycomb lattice,
 and can be viewed as the basic building block for many carbon-based materials with other dimensionalities,
 including fullerene, nanotube and graphite, etc. Its low energy band structure consists of two inequivalent
 pairs of cones with apices located at the Brillouin   zone corners\cite{mcclure}. In these cones, the energy-dispersion
 relation is linear, and the dynamics of the charge carriers is governed by  a  massless Dirac-type equation.
 The form of the electronic band structure is  expected to lead to a number of unusual electronic properties
 in graphene such as the anomalous quantized Hall effect, the absence of the weak localization and
 the existence of the minimal conductivity\cite{geim}. Graphene is also an interesting candidate
 for transport-applications, in particular for spintronics:The mobility is remarkably high  and the carrier
 density is controllable by agate voltage. In addition,  spin-dependent interactions can be exploited for
 the control of  the magnetoconductance \cite{zhang}\cite{berger}\cite{huertas}\cite{kane1}\cite{yao}\cite{ding}.
 Motivated by these facts, the spin-dependent properties of graphene are
 in the focus of current research; e.g., E. W. Hill \emph{et al.} fabricated graphene
 spin-valve device and observed a $10\%$ change in the resistance as the electrodes switch from a parallel to
 an antiparallel state\cite{hill2006ieee}.
 Recent experiments on spin injection in a single layer graphene
 show a rather long spin-flip length($\approx 1\mu m$) at room temperature\cite{tombros2007nature}.
 Spin injection into a graphene thin film has been successfully demonstrated
 by using non-local magnetoresistance
 measurements\cite{tombros2007nature}\cite{cho2007apl}\cite{ohishi2007apl}. W.H. Wang \emph{et al.}
 measured the magnetoresistance of mesoscopic graphite spin valve devices and observed a cusp-like
 feature of the magnetoresistance versus the applied bias and pointed out the importance of spin-dependent
 interfacial resistance for spin injection\cite{wang2008prb}.

 In this work we  investigate theoretically the
 spin-dependent transport through a graphene spin valve device with ferromagnetic
 leads having arbitrary spin-polarization directions. No gate voltage is applied. Utilizing
 Keldysh's nonequilibrium Green's function method\cite{haug} we calculate the density of state (DOS) and
 the electrical current in the ferromagnet-graphene-feromagnet(FM-G-FM) coupled system. The differential
 conductance and the tunnel magnetoresistance are also calculated without and with a static external magnetic
 field at finite temperatures. We found that at zero magnetic fields, the current-voltage curves in this
 spintronic structure show a nonlinear characteristic, the differential conductance as a function of the
 applied voltage exhibits  a strong dip near zero bias. The behaviour of the zero-bias anomaly in TMR is governed by
 the leads spin polarization.  With increasing the temperature,  the dip in the differential conductance and
 the cusp in TMR near zero bias diminish. When both the spin-polarization degrees and the relative angles of
 the two ferromagnetic moments are large, the differential conductance is small due to the influence of both
 the DOS in graphene and the conventional spin-valve effect. In the presence of a static magnetic field, the
 differential conductance and TMR show periodic oscillations due to a resonant transport though the Landau
 levels when the bias-voltage values are appropriate. At zero bias voltage, the differential conductance versus
 the temperature show a  behavior different from the field-free case. We attribute this fact to  the breaking of
 the insulator-type properties of graphene at finite magnetic fields.

 The rest of this paper is organized as
 follows: In section 2, we introduce the model and derive the current formula in the absence of the magnetic field.
 In section 3, the magnetotransport properties of this system are computed at finite external magnetic fields.
 In section 4, the corresponding numerical results are given. Finally, a summary is presented.%

 \section{Theoretical model}
\begin{figure}[h]
    \includegraphics[width=0.8\columnwidth ]{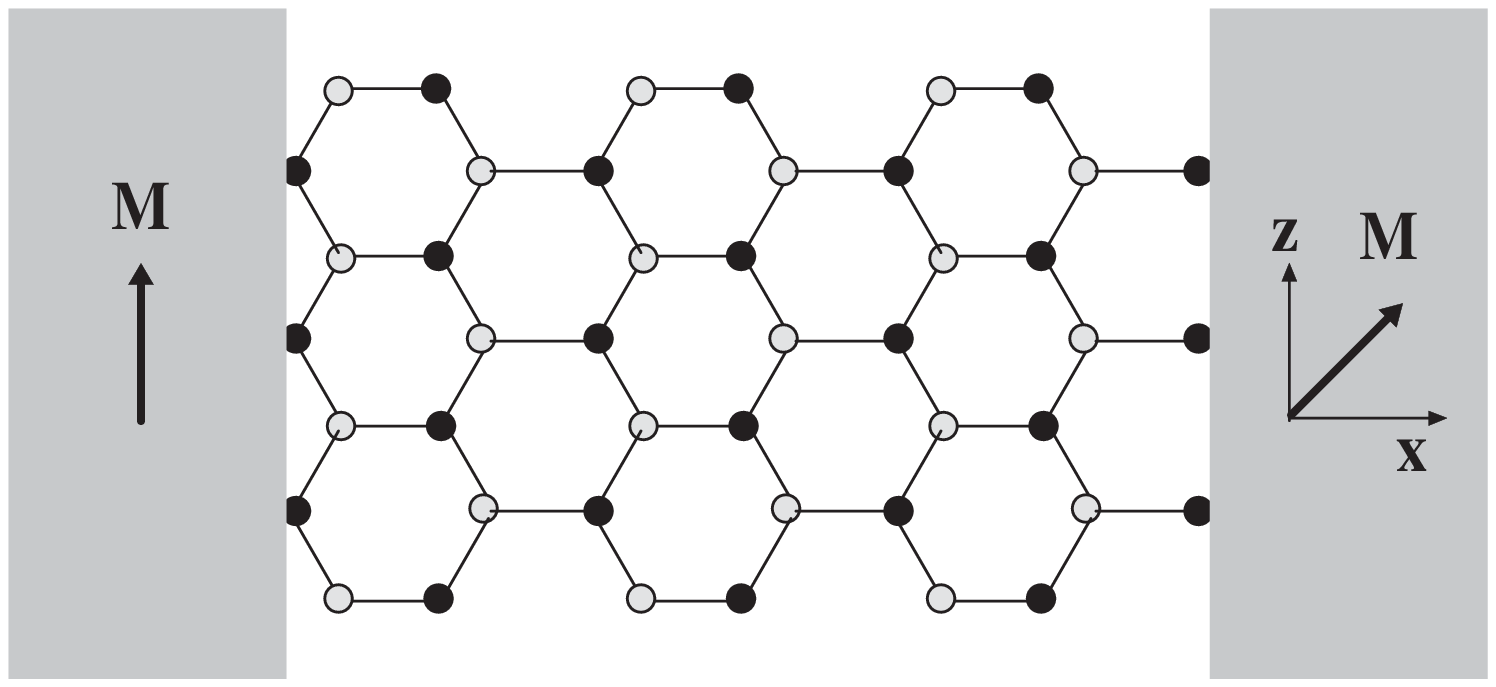}
    \caption{ Schematic illustration of the system considered in this work.
    the graphene is connected to two magnetic leads by the tunneling barriers.
    The moments of the leads are aligned by a relative angle $\theta $,
    and the coupling matrix between $\alpha(\alpha =L,R)$ electrode and graphene
    is $T_{k\alpha }$.}\label{fig1}%
    \end{figure}

 We  consider a  spin valve device consisting of a graphene layer contacted to  ferromagnetic electrodes,
 as shown in Fig.\ref{fig1}. The moment $\mathbf{M}_L$ of the left electrode is assumed to define  the $z$ direction,
 while the moment $\mathbf{M}_R$ of the right electrode deviates from the $z$ direction by a relative angle $\theta$.
 A bias voltage $V$ is applied between the left and the right electrode. The electrical current flows in the $x$
 direction. The left and the right electrodes can be described by Hamiltonians
 \begin{equation}
 H_L=\sum\limits_{\mathbf{k},\sigma}\varepsilon_{\mathbf{k}L\sigma}
 c_{\mathbf{k}L\sigma}^\dag c_{\mathbf{k}L\sigma},\end{equation}\begin{equation}H_{R}
 =\sum\limits_{\mathbf{k},\sigma }
 [\varepsilon _{R}(\mathbf{k})-\sigma \mathbf{M}_{R}\cos \theta ]c_{\mathbf{k}R\sigma }^{\dag}
 c_{\mathbf{k}R\sigma }-\mathbf{M}_{R}\sin \theta c_{\mathbf{k}R\sigma}^{\dag }c_{\mathbf{k}R\overline{\sigma }},
 \label{03}
 \end{equation}
 where $\varepsilon_{\mathbf{k}\alpha\sigma}$ is
 the single electron energy associated with the momentum $\mathbf{k}$, the spin $\sigma$
 and the $\alpha$ electrode. $c_{\mathbf{k}\alpha\sigma}^\dag(c_{\mathbf{k}\alpha\sigma})$
 creates (annihilates) an electron with the energy  $\varepsilon_{\mathbf{k}\alpha\sigma}$.

 The tight-binding Hamiltonian of the electrons in graphene is given by
 \begin{equation}
 H_G=-t\sum\limits_{\langle i,j\rangle,\sigma} (a_{i,\sigma}^\dag b_{j,\sigma}+\text{H.c.}),
 \label{hg}
 \end{equation}
 where $a_{i,\sigma}^\dag (a_{i,\sigma})$ creates (annihilates) an electron
 with the spin $\sigma$ on the position $\mathbf{R}_i$
 of the sublattice A, $b_{i,\sigma}^\dag (b_{i,\sigma})$ creates (annihilates) an electron with the spin
 $\sigma$ on the position $\mathbf{R}_i$ on the sublattice B, and $t$ is
 the nearest neighbor $(\langle i,j\rangle)$ hopping energy. In the momentum space
 the Hamiltonian $H_G$ is rewritten as
 \begin{equation}
 H_G=\sum\limits_{\mathbf{q},\sigma}[\phi(\mathbf{q})a_{\mathbf{q}\sigma}^\dag b_{\mathbf{q}\sigma}
 +\phi(\mathbf{q})^* b_{\mathbf{q}\sigma}^\dag a_{\mathbf{q}\sigma} ],\label{hg12}
 \end{equation}
 where
 \[ \phi(\mathbf{q})=-t\sum\limits_{i=1}^3e^{i\mathbf{q}\cdot\mathbf{\delta_i}}
 \quad \mbox{ with}\quad\delta_1=\frac{a}{2}(1,\sqrt{3},0),\;\delta_2=\frac{a}{2}(1,-\sqrt{3},0),\;
 \delta_3=a(1,0,0).\]
 Here $a$ is the lattice spacing.
 Diagonalizing the  Hamiltonian (\ref{hg12})
 one finds
 \[ E_{\pm}(\mathbf{k})=\pm t|\phi(\mathbf{k})|,\]
 which can be linearized around the $\mathbf{K}$
 points of the Brillouin zone
 leading to the dispersion
 \begin{equation}
 E_{\pm}(\mathbf{k})=\pm v_F|\mathbf{k}|,
 \end{equation}
 where $v_F=3ta/2$ is the Fermi velocity of electron ($t\sim 2.3eV$ \cite{Gusynin}) .
 The coupling between the electrodes and graphene is modeled by
 \begin{equation}
 H_T=\frac{1}{\sqrt{N}}\sum\limits_{\mathbf{kq}\alpha\sigma}
 [T_{\mathbf{k}\alpha\mathbf{q}}c_{\mathbf{k}\alpha\sigma}^\dag a_{\mathbf{q}\sigma}
 + \text{H.c.}], \ \ \alpha=L,R.
 \end{equation}
 $T_{\mathbf{k}\alpha\mathbf{q}}$ is the coupling matrix between the $\alpha$ electrode and the graphene;
 $N$ is the number of sites on the sublattice A.

 The electrical current from the left
 electrode to the graphene sheet is obtained from the time evolution of the occupation
 number operator of the left electrode, i.e.
 \begin{equation}
 I=e\langle\dot{ \hat{\mathcal{N}_L}}\rangle
 =\frac{ie}{\hbar}\langle[H,\mathcal{\hat N}_L]\rangle,\quad \mathcal{\hat N}_L
 =\sum\limits_{\mathbf{k}\sigma}c_{\mathbf{k}L\sigma}^\dag c_{\mathbf{k}L\sigma}.
 \label{jl}
 \end{equation}
 Using the nonequilibrium Green's function method,
 Eq.(\ref{jl}) can be further
 expressed as
 \begin{equation}
 \begin{array}{cll}
 I &=&-\frac{ie}{\hbar N}\int\frac{d\varepsilon}{2\pi}
 Tr\sum\limits_{\mathbf{qq'}}\left\{[G_{\mathbf{q}a,\mathbf{q'}a}^{r}(\varepsilon)
 -G_{\mathbf{q}a,\mathbf{q'}a}^{a}(\varepsilon)]f_L(\varepsilon)+G_{\mathbf{q}a,\mathbf{q'}a}^{<}(\varepsilon)\right\}
 \Gamma_{L\mathbf{q'q}}(\varepsilon),
 \end{array}\label{jl3}
 \end{equation}
 where $Tr$ means the trace in the spin space and $f_\alpha(\varepsilon)$ is the Fermi distribution function
 at the energy $\varepsilon$.
 \[ G_{\mathbf{q}a,\mathbf{q'}a}^{\sigma\sigma',<}(t-t')=i\langle a_{q'\sigma'}^\dag(t') a_{q\sigma}(t)\rangle\]
 is the matrix expression for the lesser Green's function.
 $G_{\mathbf{q}a,\mathbf{q'}a}^{r}(\varepsilon) $ and $G_{\mathbf{q}a,\mathbf{q'}a}^{a}(\varepsilon)$
 are  $2\times 2$ matrices  in the spin space in the sublattice A describing respectively
 the retarded and the advanced Green's function. The line width matrix $\Gamma_{\alpha \mathbf{qq'}}$
 is given by
 \begin{equation}
 \Gamma_{\alpha \mathbf{qq'}}(\varepsilon)=\left(\begin{array}{cc}
 \Gamma_{\alpha \mathbf{qq'}}^\uparrow & 0 \\0 & \Gamma_{\alpha \mathbf{qq'}}^\downarrow%
 \end{array}%
 \right),
 \quad \mbox{and}\quad \Gamma_{\alpha\mathbf{qq'}}^\sigma(\varepsilon)
 =2\pi\sum\limits_\mathbf{k}T_{\mathbf{k}\alpha\mathbf{q}}^*T_{\mathbf{k}\alpha\mathbf{q'}}
 \delta(\varepsilon-\varepsilon_{\mathbf{k}\alpha\sigma}),
 \end{equation}
 where $T_{\mathbf{k}L\mathbf{q}}$
 stand for the  coupling of  graphene to the electrodes.
  To evaluate $I$ from Eq.(\ref{jl3}) the retarded  Green's function $G_{\mathbf{q}a,\mathbf{q'}a}^{r}
 (\varepsilon)$ needs to be calculated.
 Here we consider electrons near the Fermi level which contribute
 predominantly  to tunneling. In this case one may assume
 the coupling matrix $T_{\mathbf{k}L\mathbf{q}}$ to be independent of $\mathbf{q}$
 and set $\Gamma_{\alpha\mathbf{qq'}}^\sigma=\Gamma_{\alpha}^\sigma$. Standard Green's function
 technique \cite{haug} delivers
 \begin{equation}
 \begin{array}{cll}
 G_{qa,q'a}^{r}(\varepsilon)&=&\delta_{qq'}g_{qa,qa}^{r}(\varepsilon)+ g_{qa,qa}^{r}(\varepsilon)
 T(\varepsilon)g_{q'a,q'a}^{r}(\varepsilon).
 \end{array}\label{aa3}
 \end{equation}
 In the above expressions we introduced
 \[ g_{qa,qa}^{r,a}(\varepsilon)=\frac{\varepsilon}{(\varepsilon\pm i\eta)^2-|\phi(q)|^2},
 \quad \mbox{ and}\quad T(\varepsilon)=\frac{\Sigma^r(\varepsilon)}{1-\overline{g}_{aa}^{r}(\varepsilon)
 \Sigma^r(\varepsilon)},\]
 where
 \[ \overline{g}_{aa}^{r,a}(\varepsilon)
 =\frac{1}{N}\sum\limits_\mathbf{q}g_{\mathbf{q}a,\mathbf{q}a}^{r,a}(\varepsilon),
 \quad \mbox{ and}\quad\Sigma^{r,a}(\varepsilon)=\mp\frac{i}{2}[\Gamma_{L}(\varepsilon)+R
 \Gamma_{R}(\varepsilon)R^\dag]\]
 with
 $$R=\left(
 \begin{array}{cc}
 \cos \frac{\theta }{2} & -\sin \frac{\theta }{2} \\
 \sin \frac{\theta }{2} & \cos \frac{\theta }{2}
 \end{array}
 \right) .
 $$In Eq.(\ref{jl3}),
 $G_{\mathbf{q}a,\mathbf{q'}a}^{<}(\varepsilon)$ can be derived by applying Keldysh equation
 $$G_{\mathbf{q}a,\mathbf{q'}a}^{<}(\varepsilon)
 =G_{\mathbf{q}a,\mathbf{q'}a}^{r}(\varepsilon)\Sigma^<(\varepsilon)G_{\mathbf{q}a,\mathbf{q'}a}^{a}(\varepsilon)$$
 with
 $$\Sigma^<(\varepsilon)=i[f_L(\varepsilon)\Gamma_{L}(\varepsilon)
 +f_R(\varepsilon)R \Gamma_{R}(\varepsilon)R^\dag ].$$
  Substituting the expressions of the graphene Green's functions in Eq.(\ref{jl3}),
  and after a straightforward calculation we obtain the tunneling current as
  \begin{equation}
  \begin{array}{cll}
  I&=&\frac{e}{\hbar}\int\frac{d\varepsilon}{2\pi}
  \text{Tr}\{[\mathcal{G}_{a}^{r}(\varepsilon)(R\Gamma_R(\varepsilon)R^\dag)]
  \mathcal{G}_{a}^{a}(\varepsilon)\Gamma_{L}(\varepsilon)\}[f_R(\varepsilon)-f_L(\varepsilon)],\\
  \end{array} \label{fj}
  \end{equation}
  where
  \begin{equation}
  \mathcal{G}^{r,a}_a(\varepsilon)
  =\sum\limits_{q'}\sum\limits_qG_{qa,q'a}^{r,a}(\varepsilon)
  =\frac{\overline{g}_{aa}^{r,a}(\varepsilon)}{1-\overline{g}_{aa}^{r,a}(\varepsilon)
  \Sigma^{r,a}(\varepsilon)}.\label{sumaa2}
  \end{equation}
  Introducing a cutoff $k_c$ leads to
  \begin{equation}\overline{g}_{aa}^{r,a}(\varepsilon)
  =-F_0(\varepsilon)\mp i\pi\rho_0(\varepsilon),\label{gaa1}
  \end{equation}
  where
  \begin{equation}
  F_0(\varepsilon)
  =\frac{\varepsilon}{D^2}\ln\frac{|\varepsilon^2-D^2|}{\varepsilon^2},\ \
  \rho_0(\varepsilon)=\frac{|\varepsilon|}{D^2}\theta(D-|\varepsilon|).
  \end{equation}
  $D=v_Fk_c$ stands for a high-energy cutoff of the graphene  bandwidth.
  Invoking the Debye's prescription, we choose $k_c$ such  that the total number of states in the Brillouin zone
  is  conserved after the linearization of the spectrum around the $\mathbf{K}$  point.
  Hence, Eq.(\ref{gaa1}) is accurate for $\varepsilon\ll D$, i.e.
  $\varepsilon$ has to be in the region where  the linearization of the spectrum is justifiable
  which   is roughly estimated   \cite{castro} to be [-1.6 eV, 1.6 eV].
  In Eq.(\ref{sumaa2})
  we assumed a symmetrical voltage drop as $\mu_{L,R}=E_F\pm\frac{1}{2}eV$,
  and put $E_F=0$ in the numerical calculations.
  The TMR at the angle $\theta$ is conventionally
  defined as
  \begin{equation}
  TMR(\theta)=\frac{I(0)-I(\theta)}{I(0)}.
  \end{equation}

  \section{Magnetotransport at finite external magnetic fields}
  In the presence of a  static external magnetic field, the description of
  the transport properties of electrons in a honeycomb lattice becomes much more involved due to the
  coupling between graphene and the electrodes associated with the Hofstadter problem.
  To circumvent this situation we describe the electrons in the honeycomb lattice as Dirac fermions in the continuum.
  At first we introduce  the field operators\cite{peres2006prb}\cite{dora2007prb}
  \begin{equation}
  \Psi_\sigma(\mathbf{r})=\sum\limits_q\frac{e^{iqx}}{\sqrt{L}}
  \left(\begin{array}{c}0\\\phi_0(y)\end{array}\right)d_{q\sigma}+\sum\limits_{q,n,\alpha}
  \frac{e^{iqx}}{\sqrt{2L}}\left(\begin{array}{c}\phi_n(y-ql_B^2)\\\phi_{n+1}(y-ql_B^2)
  \end{array}\right)d_{qn\alpha\sigma},
  \end{equation}
  where $l_B=1/\sqrt{eB}$ is the cyclotron length, $n=0,1,2,...$, $\alpha=\pm 1$,
  and $\phi_n(x)$ is the nth eigenfunction of the usual one-dimensional harmonic oscillator.
  The Hamiltonian describing  the electrons in graphene acquires the second-quantized form %
  \begin{equation}
  H_G=\sum\limits_{qn\alpha\sigma}E(n,\alpha)d_{qn\alpha\sigma}^\dag d_{qn\alpha\sigma}\label{hg2}
  \end{equation}
  where
  $$E(n,\alpha)=\alpha\omega_c\sqrt{n+1}$$
  is the Landau level with $\omega_c=v_F\sqrt{2eB}$. The sum over integer $n$'s is cut off at $\mathcal{N}$.
  The coupling between graphene and ferromagnetic electrodes is
  \begin{equation}
  \begin{array}{lll}
  H_T&=&\sum\limits_{kq\sigma}\phi_0(-ql_B^2)(T_{kLq}c_{kL\sigma}^\dag d_{q\sigma}
  +T_{kLq}^* d_{q\sigma}^\dag c_{kL\sigma})\\
  &&+\sum\limits_{kqn\alpha\sigma}[\phi_n(-ql_B^2)+\alpha\phi_{n+1}(-ql_B^2)]
  [T_{kLq}c_{kL\sigma}^\dag d_{qn\alpha\sigma}+T_{kLq}^* d_{qn\alpha\sigma}^\dag c_{kL\sigma}]\\
  &&+ \sum\limits_{kq\sigma}\phi_0(L-ql_B^2)(T_{kRq}c_{kR\sigma}^\dag d_{q\sigma}+T_{kRq}^* d_{q\sigma}^\dag
  c_{kR\sigma})\\
  &&+\sum\limits_{kqn\alpha\sigma}[\phi_n(L-ql_B^2)+\alpha\phi_{n+1}(L-ql_B^2)][T_{kRq} c_{kR\sigma}^\dag
  d_{qn\alpha\sigma}+T_{kRq}^* d_{qn\alpha\sigma}^\dag c_{kR\sigma}].
  \end{array}
  \end{equation}
  where $L$ is length of graphene. Similar to the calculation of Eq.(\ref{fj})
  we obtain the electrical current in the form
  \begin{equation}
  \begin{array}{cll}
  J_L&=&\frac{e}{\hbar}\int\frac{d\varepsilon}{2\pi}Tr\{[\varepsilon-(1+\varepsilon X)
  \Sigma^r ]^{-1}(1+\varepsilon X)^2R\Gamma_{R}R^\dag[\varepsilon-(1+\varepsilon X)\Sigma^a]^{-1}
  \Gamma_L\}[f_R(\varepsilon)-f_L(\varepsilon)],
  \end{array}
  \end{equation}
  where
  $$X(\varepsilon)=\sum\limits_{n\alpha}\frac{1}{\varepsilon-E(n,\alpha)}
  =\frac{2\varepsilon}{\omega_c^2}[\Psi(\frac{\omega_c^2-\varepsilon^2}{\omega_c^2})
  -\Psi(\frac{(\mathcal{N}+2)\omega_c^2-\varepsilon^2}{\omega_c^2})]
  $$
  with $\Psi(z)$ denoting the digamma function \cite{abra}.
  The electric conductance can be obtained by $\partial J_L/\partial V$.
  For a small bias voltage we obtain the Landauer-B\"uttiker-type expression
  \begin{equation}
  G=\frac{e^2}{h}T_{eff},
  \end{equation}
  where
  $$T_{eff}=Tr\{[E_F-(1+E_F X(E_F))
  \Sigma^r ]^{-1}(1+E_FX(E_F))^2R\Gamma_{R}R^\dag[E_F-(1+E_F X(E_F))\Sigma^a]^{-1}\Gamma_L\}
  $$
   plays the role of  an effective energy-dependent transmission coefficient.
   In the limit $B\rightarrow0$, we further obtain for $\theta=0$,
   \begin{equation}
   T_{eff}=\sum\limits_\sigma \frac{\lambda\Gamma_L^\sigma\Gamma_R^\sigma}{E_F^2+\frac{1}{4}
   \lambda(\Gamma_L^\sigma+\Gamma_R^\sigma)^2},
   \end{equation}
   where
   $ \lambda=(2\mathcal{N}+3)^2+2(\mathcal{N}+1)(\mathcal{N}+2)(2\mathcal{N}+3)\frac{\omega_c^2}{E_F^2}.$

   \section{Numerical results and discussions}

\begin{figure}[h]
    \includegraphics[width=0.8\columnwidth ]{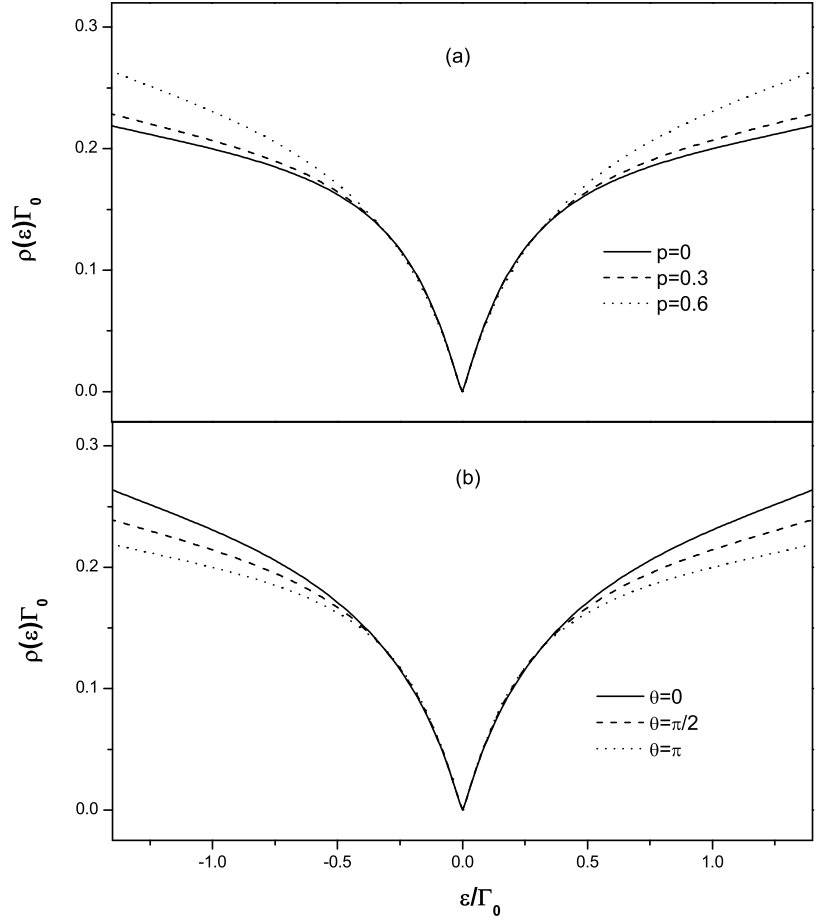}
    \caption{DOS as a function of energy $\varepsilon$ for
    different polarization $p$ at $\theta=0$ (a) and for different angle $\theta$ at $p=0.6$ (b).
    The other parameters are taken as$D=2\Gamma_0$ and $B=0$. }\label{fig2}%
    \end{figure}

   Adopting the wide bandwidth approximation for the graphene spin valve system we neglect the energy dependence
   of the line width functions $\Gamma_\alpha^\sigma(\varepsilon)$.  Denoting the spin polarization of
   the left and the right electrodes by respectively $p_L$ and $p_R$
   we write $\Gamma _{L}^{\uparrow ,\downarrow }=\Gamma _{R}^{\uparrow,\downarrow }
   =\Gamma _{0}(1\pm p)$, here $\Gamma_0$ describes the coupling between the graphene and the
   electrode without the internal magnetization. Here we  assumed  the left and the
   right electrodes to be of the same material. In the following numerical calculation,
   we take $\Gamma_0$ as the energy scale. We calculate the DOS in graphene
   via relation $\rho(\varepsilon)=-\frac{1}{\pi}\text{Im}\sum\limits_\sigma\mathcal{G}_a^{\sigma\sigma,r}(\varepsilon)$.
   Fig.\ref{fig2} shows the DOS as a function of energy for different polarization $p$ and magnetization
   angle $\theta$. It is clearly observed that the DOS displays a dip structure with the energy. For nonzero
   energy, the DOS increases with increasing $p$, however decreases with increasing $\theta$. This is
   caused by the different tunneling rates for up and down spins owing to the splitting of DOS of the
   ferromagnetic leads. This splitting acts as an effective magnetic field\cite{zhu2008pla} reaching
   values much larger than externally applied magnetic field\cite{Pasupathy2004sci}. This results
   in a spin dependence of the DOS in the central region. While at the zero energy point, the DOS is
   independent of $p$ and $\theta$. This stems from the nature of Dirac point in the graphene.

\begin{figure}[h]
    \includegraphics[width=0.8\columnwidth ]{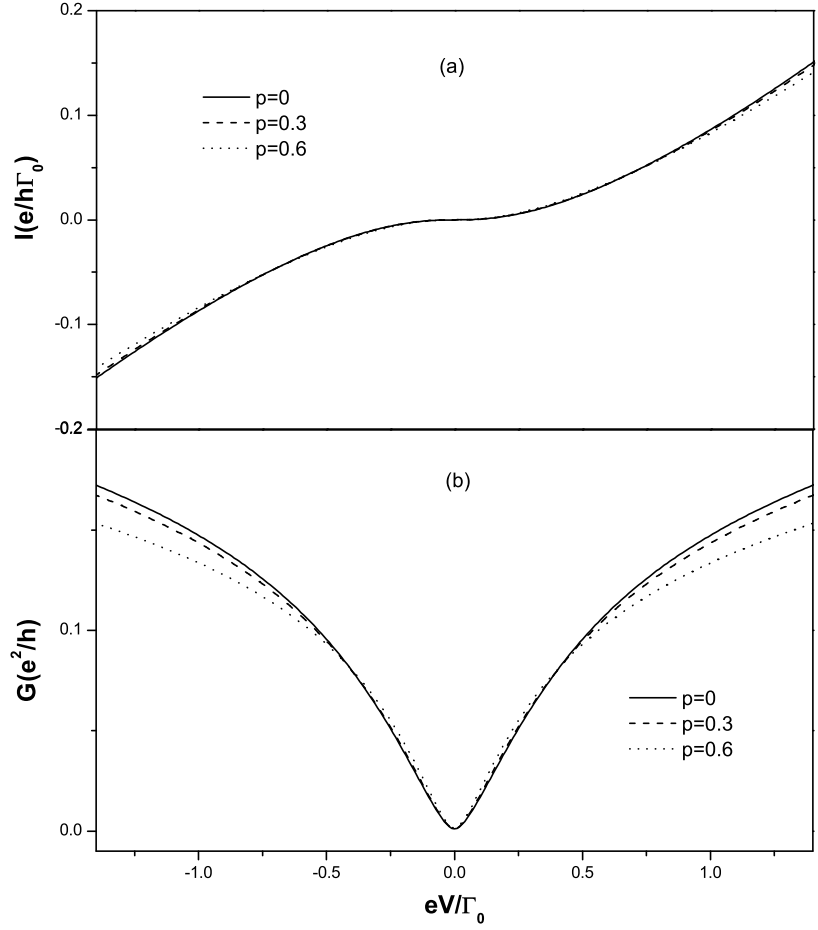}
    \caption{ The bias dependence of the electrical current $I$
    (a) and differential conductance $G$ (b) for different polarization $p$ at $\theta=0$
    and $k_BT=0.005\Gamma_0$. The other parameters are taken the same as those of Fig.\ref{fig2}. }\label{fig3}
    \end{figure}

   The bias dependence of the electrical current and the differential
   conductance $G=dI/dV$ are  shown
   in Fig.\ref{fig3} for  parallel electrodes  magnetizations  and   for the different polarization $p$,
   The nonlinear behaviour of $G$ and the strong dip at zero
   bias are in line with the experimental observations \cite{wang2008prb}, and are at variance
   with the typical behaviour when a Fermi liquid were in  central region instead of graphene.
   In that case the electrical current is proportional to the applied voltage at small bias
   and obey the Ohmic law\cite{mu2005prb}. The results obtained here resemble rather the
   ferromagnet-insulator-ferromagnet(FM-I-FM) junctions\cite{mooderaprl1995}\cite{mooderaprl1998}.
    The reason is that: the DOS in graphene diminishes at the Fermi level. Hence, graphene
    sheet can be viewed as a  tunneling barrier at the zero energy point, similar to FM-I-FM system. %
    With increasing the polarization $p$, the portion of spin-up electron states increases while for
    the spin-down electrons decreases. However, the scattering of the former is larger than that of
    the latter, thus we conclude that the conductance $G$
    decreases with increasing $p$ at nonzero bias (cf. Fig.\ref{fig3} (b)).
    While $G$ is almost independent of $p$ at zero bias, which stems from the fact that the spin
    transport through the Dirac point of graphene is ballistic due to its
    insulator-like properties.

\begin{figure}[h]\includegraphics[width=0.8\columnwidth ]{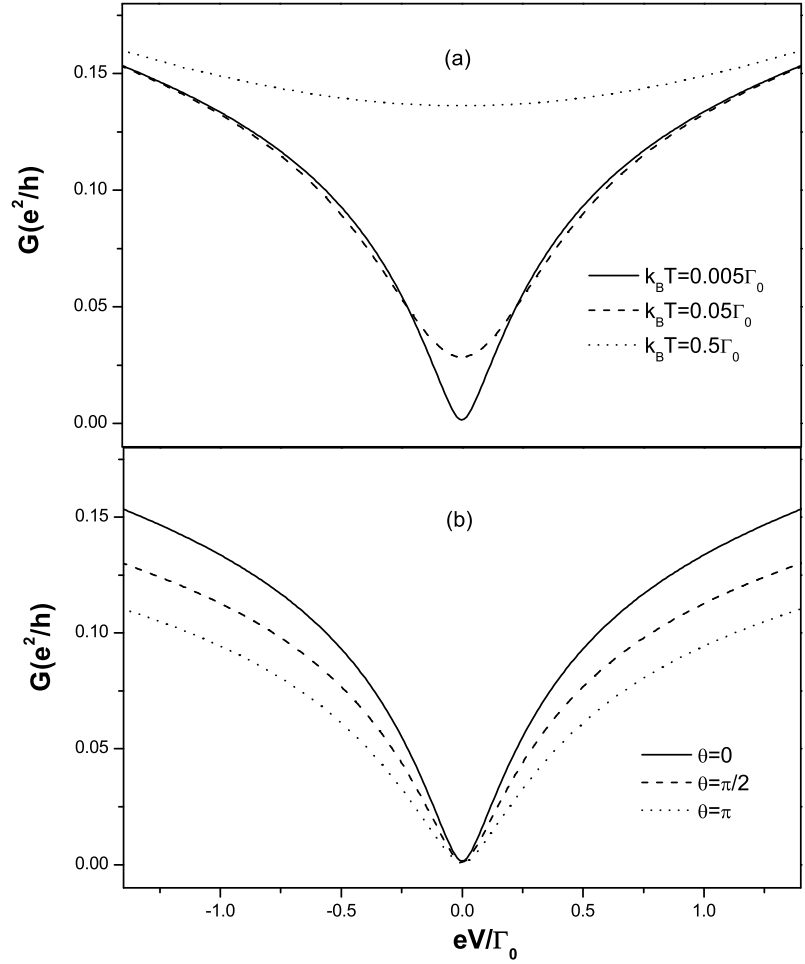}
    \caption{The bias dependence of the differential conductance $G$ for different temperature $T$
    at $\theta=0$ (a) and for different angle $\theta$ at $k_BT=0.005\Gamma_0$ (b).
    The parameters are taken the same as those of Fig.\ref{fig2}.}\label{fig4}%
    \end{figure}

\begin{figure}[h]\includegraphics[width=0.8\columnwidth ]{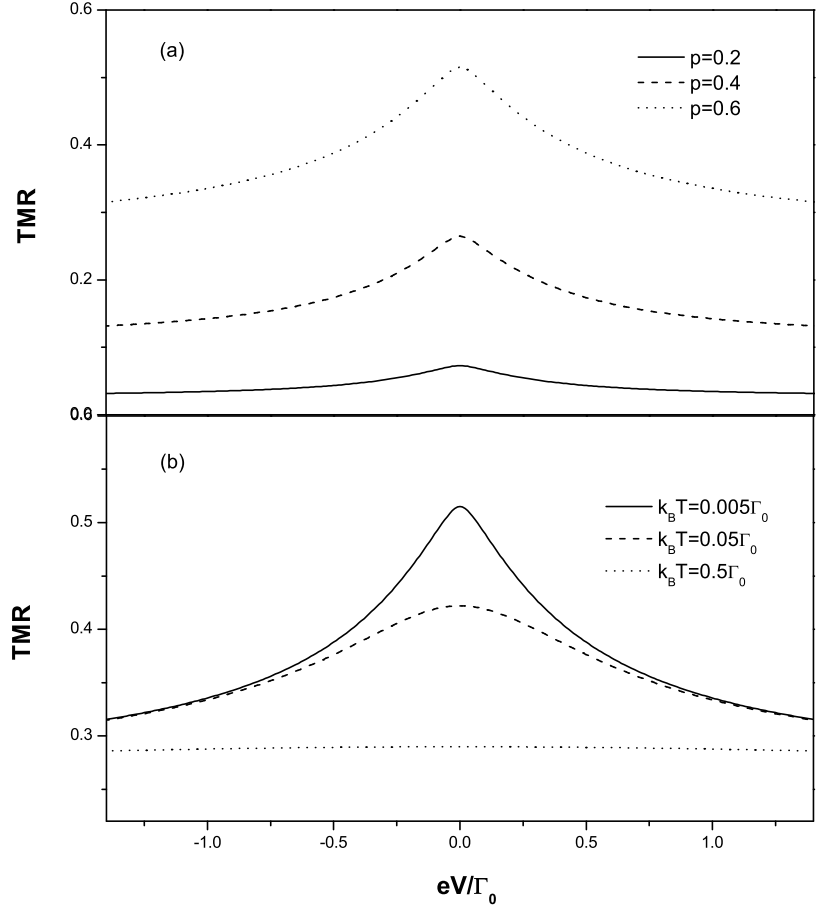}
    \caption{The bias dependence of TMR for different polarization $p$ at $k_BT=0.005\Gamma_0$
    (a) and for different temperature $T$ at $p=0.6$ (b).
    The other parameters are taken the same as those of Fig.\ref{fig2}. }\label{fig5}%
    \end{figure}

    The bias dependence of the conductance at different temperatures $T$
    and angles $\theta$ are shown in Fig.\ref{fig4}: The conductance is roughly  independent of $T$
    at large bias. Near $V=0$ the conductance increases with increasing $T$. This behavior is also in
    contrast to  usual Fermi liquids  where the conductance decreases with increasing temperature because
    thermal fluctuations enhance  the scattering of conduction electrons and thereby contributes to the
    resistance of system \cite{mu2005prb}. In our case, the graphene is equivalent to a barrier at the
    zero energy point. Near zero bias voltage, the thermally excited electrons are dominant in tunneling
    process. Therefore, with increasing temperatures, the increase of the thermally excited electrons
    enhances the conductance. On the other hand for large bias, the contributions to the conductance
    stem mainly  from  electrons with excess energies well above the Fermi level (as dictated by the
    applied voltage) which leads to a very weak dependence of the conductance on
    the temperature at large bias.
    The conductance  as a function of  $\theta$  (Fig.\ref{fig4} (b)) follows the
    conventional  behaviour of magnetic junctions such as  the ferromagnet-quantum dot-ferromagnet
    system \cite{sergueevprb2002}\cite{muprb2006}\cite{zzgprb2004}. When $\theta $ changes from $0$
    to $\pi $ the number of spin-up and spin-down electrons is rearranged. Fig.\ref{fig5} shows
    TMR ratio as a function of the applied voltage for different polarizations $p$ and temperatures $T$.
    A pronounced cusp-like feature appears at zero bias in line with experimental observations\cite{wang2008prb}.
    We  assign this behaviour  to  the result of a  non-trivial combined effect of graphene and conventional
    spin-valve properties, evidenced by the dependence of TMR on the polarization $p$ at
    a fixed bias voltage (cf. Fig.\ref{fig5}(a)). When increasing the polarization $p$
    the contribution of spin-up states relative to the spin-down  is increased resulting
    in an increase of TMR for the entire bias range. However, the TMR changes in a
    non-linear manner: the TMR value at zero bias becomes larger than that at nonzero bias.
    This is because for a small bias graphene behaves as an insulator. In contrast, for a
    large bias graphene behaves more like a metal, which is an essential difference between
    graphene and other materials in the central region. The spin tunneling is ballistic through
    an insulator in contrast to a metal, whence the TMR is more enhanced around zero bias.
    There is almost no changes with the temperature $T$ for large bias (see Fig.\ref{fig5} (b)).
    With increasing temperatures more thermally excited electrons contribute to the electrical
    currents for the parallel and the antiparallel configurations, in fact, the increase of
    the electrical current is faster for the antiparallel configuration. Therefore,
    we can conclude that the TMR decreases at higher temperatures and the zero-bias
    anomaly diminishes in this situation.

\begin{figure}[h]\includegraphics[width=0.8\columnwidth ]{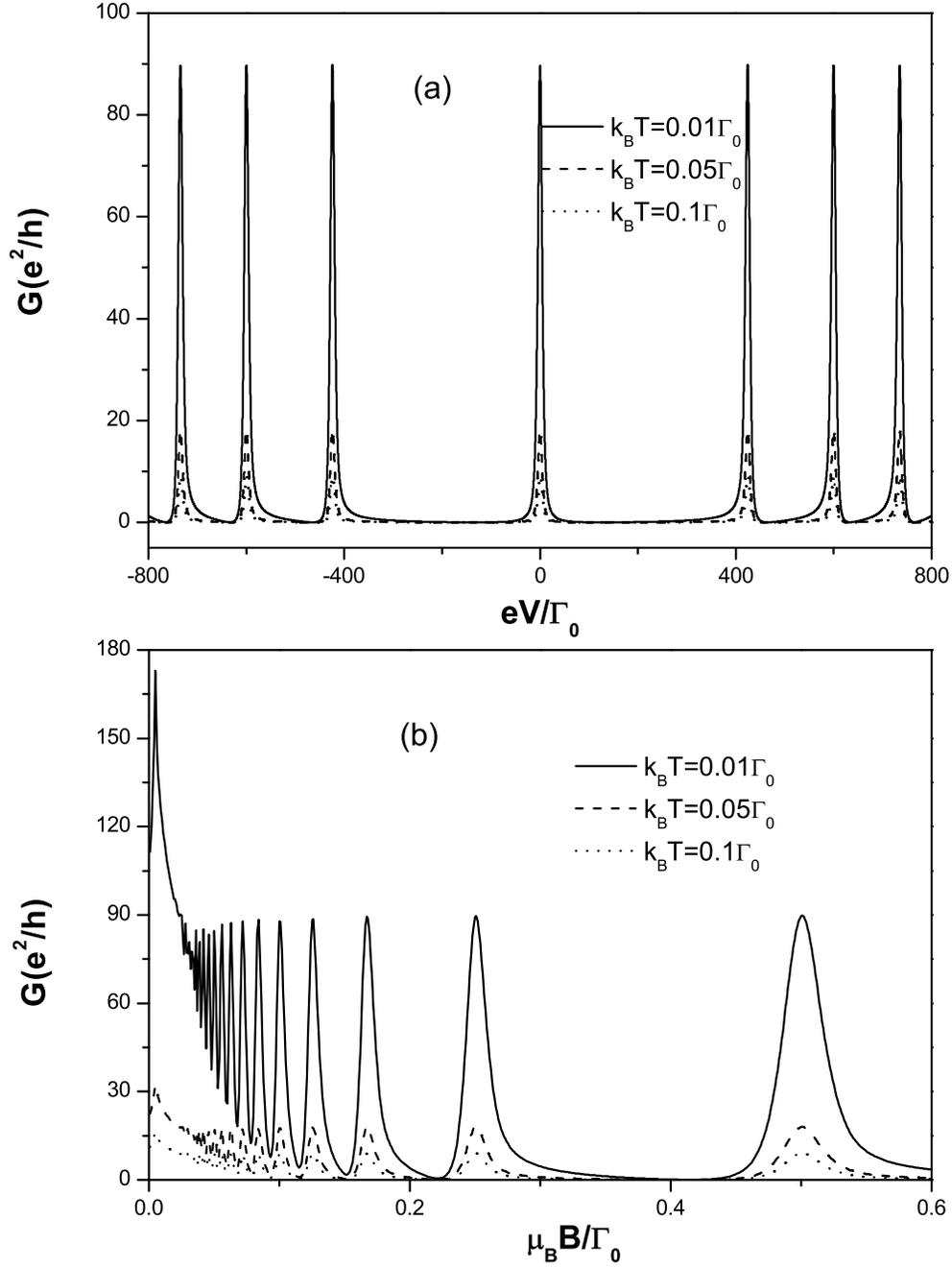}
    \caption{The differential conductance as a function of the bias voltage at $\mu_BB=1\Gamma_0$ (a)
    and of the magnetic field at $eV=300\Gamma_0$ (b) for different temperature $T$
    in the parallel configuration. The parameters are taken as $p=0.6$ and $\mathcal{N}=100$. }\label{fig6}%
    \end{figure}

\begin{figure}[h]\includegraphics[width=0.8\columnwidth ]{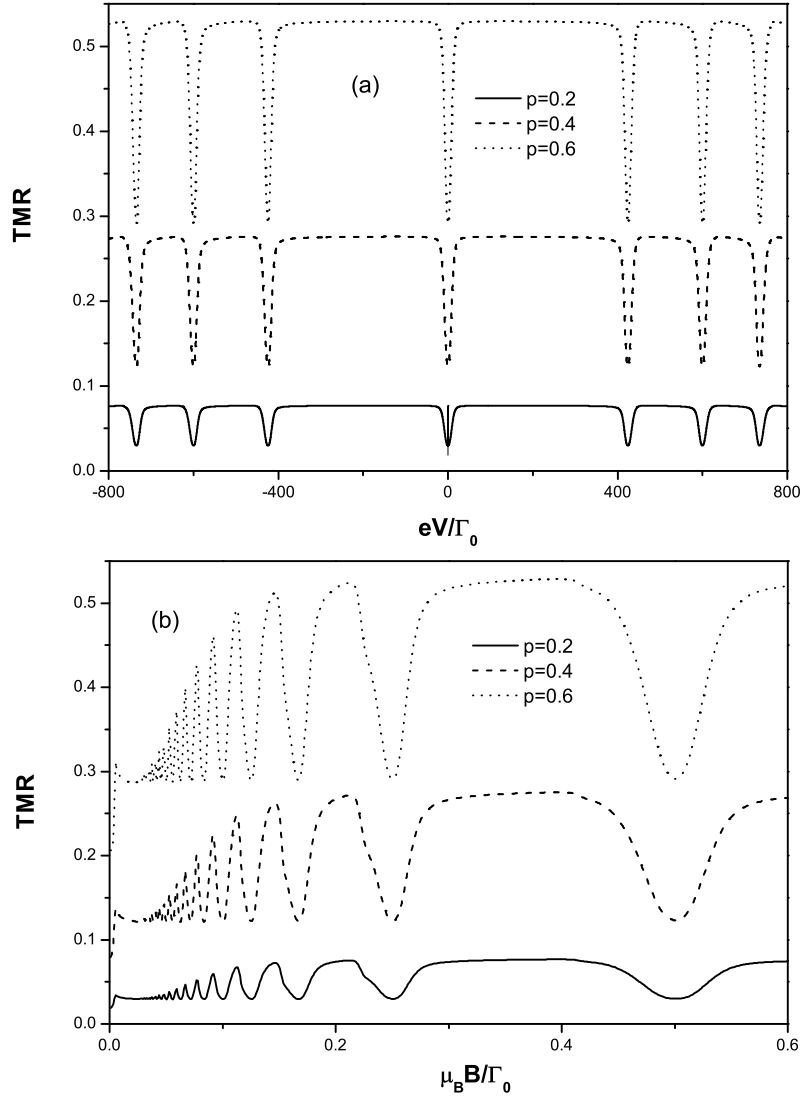}
    \caption{The TMR as a function of the bias voltage at $\mu_BB=1\Gamma_0$ (a) and of
    the magnetic field at $eV=300\Gamma_0$ (b) for different polarization $p$ at $k_BT=0.1\Gamma_0$.
    The other parameters are taken the same as those of Fig.\ref{fig6}}\label{fig7}%
    \end{figure}

    The bias and the magnetic field dependencies
    of the differential conductance $G$ for the different temperature in the parallel configuration of
    magnetization is shown in Fig.\ref{fig6}. $G$ versus the applied voltage exhibits an oscillating
    behaviour. Each conductance peak corresponds to resonant transport through a Landau level.
    As bias voltage increases, the distance between  two neighboring  peaks decreases
    due to the decrease of the distance between neighboring Landau levels.
    The differential conductance oscillates as a function of the magnetic field,
    as shown in Fig.\ref{fig6} (b). This is due to resonant transport through the Landau
    levels at a particular magnetic field strength and the applied bias values.
    With increasing  temperatures  (Fig.\ref{fig6}(a)) the differential conductance
    deceases for all  bias voltages. In particular, at zero bias the temperature dependence
    of the differential conductance is different from the magnetic field-free case.
    The explanation for this phenomena is as follows: the magnetic field lifts the
    insulator behavior at Dirac point in graphene, and thus  thermal fluctuations suppress
    the conductance. Fig.\ref{fig7} shows the TMR as a function of the bias
    voltage and the magnetic field strength  for different polarizations p.  TMR reaches a minimal
    value at  bias voltages corresponding to the conductance peaks and increases with increasing $p$
    which is nothing but a spin-valve effect.

    \section{Summary}
    In conclusion, we have studied the
    spin-dependent transport through a graphene spin valve device for a non-collinear configuration
    by means of Keldysh's nonequilibrium Green's function method. It is found that at a zero
    magnetic field, the current-voltage curves show a nonlinear behaviour. The corresponding
    differential conductance exhibits a strong dip near zero bias.  The TMR shows a zero-bias
    anomaly that depends on the leads-spin polarization. Increasing the bias TMR follows a
    behaviour akin to a metallic or an insulating system depending on the value of the bias.
    In the presence of an external magnetic field, the differential conductance and TMR  oscillate
    periodically due to a resonant transport through Landau levels. At zero bias the differential
    conductance versus the temperature reveals a behavior different from the magnetic field-free case.

    \begin{acknowledgments}
    { The work of K.H.D. was supported by the Natural Science Foundation of Hunan Province,
    China (Grant No. 08JJ4002 ), the National
    Natural Science Foundation of China (Grant Nos. 60771059) and
    Education Department of Hunan Province,
    China.  J.B. andZ.H.Z. were supported by the cluster of excellence "Nanostructured Materials"
    of the state Saxony-Anhalt. }
    \end{acknowledgments}

    \newpage

    \end{document}